\documentclass[conference,a4paper]{APSIPA2025}
\usepackage{amsmath}
\usepackage{graphicx}
\usepackage{multirow}
\usepackage{threeparttable}
\usepackage[backend=biber,style=ieee,]{biblatex}
\usepackage{geometry}
\usepackage{fancyhdr}

\fancypagestyle{firststyle}{
  \fancyhf{}
  \fancyhead[C]{2025 Asia Pacific Signal and Information Processing Association Annual Summit and Conference (APSIPA ASC)}
}

\begin{document}

\title{Directional Selective Fixed-Filter Active Noise Control Based on a Convolutional Neural Network in Reverberant Environments}

\author{
\authorblockN{
Boxiang Wang\authorrefmark{1}, Zhengding Luo\authorrefmark{1}, Haowen Li\authorrefmark{1}, Dongyuan Shi\authorrefmark{2}, Junwei Ji\authorrefmark{1}, Ziyi Yang\authorrefmark{1}, Woon-Seng Gan\authorrefmark{1}
}

\authorblockA{
\authorrefmark{1}
School of Electrical and Electronic Engineering, Nanyang Technological University, Singapore \\
Email: \{boxiang001, luoz0021, junwei002, ziyi016\}@e.ntu.edu.sg, \{haowen.li, ewsgan\}@ntu.edu.sg}

\authorblockA{
\authorrefmark{2}
Center of Intelligent Acoustics and Immersive Communications, Northwestern Polytechnical University, China \\
E-mail: dongyuan.shi@nwpu.edu.cn
}
}

\maketitle
\thispagestyle{firststyle}
\pagestyle{fancy}

\begin{abstract}
Selective fixed-filter active noise control (SFANC) is a novel approach capable of mitigating noise with varying frequency characteristics. It offers faster response and greater computational efficiency compared to traditional adaptive algorithms. However, spatial factors, particularly the influence of the noise source location, are often overlooked. Some existing studies have explored the impact of the direction-of-arrival (DoA) of the noise source on ANC performance, but they are mostly limited to free-field conditions and do not consider the more complex indoor reverberant environments. To address this gap, this paper proposes a learning-based directional SFANC method that incorporates the DoA of the noise source in reverberant environments. In this framework, multiple reference signals are processed by a convolutional neural network (CNN) to estimate the azimuth and elevation angles of the noise source, as well as to identify the most appropriate control filter for effective noise cancellation. Compared to traditional adaptive algorithms, the proposed approach achieves superior noise reduction with shorter response times, even in the presence of reverberations.

\end{abstract}

\section{Introduction}
Active noise control (ANC) is an advanced technique that effectively attenuates low-frequency noise, offering a more compact and lightweight alternative to traditional passive noise control methods \cite{elliott1993active}. In an ANC system, a secondary source is driven by the control signal to generate anti-noise with equal amplitude but opposite phase as the unwanted noise, thereby reducing the disturbance through the sound destructive interference principle \cite{kuo1999active}. Owing to its compact size and effective noise reduction capabilities, ANC has been widely applied in various applications \cite{shen2022adaptive,jung2019local}. To address time-varying noise and dynamic acoustic environments, adaptive algorithms such as the filtered-x least mean squared (FxLMS) algorithm are commonly employed to update the control filter in real time \cite{morgan1980analysis,ji2025self}. However, these algorithms often involve high computational complexity, slow convergence speed, and are at risk of divergence \cite{wang2024computation,ji2025preventing}. To ensure simplicity and stability, many commercial ANC systems utilize pre-trained control filters with fixed coefficients \cite{shen2022adaptive}. Nevertheless, the performance of such fixed-filter approaches is highly sensitive to variations in noise source type and direction-of-arrival (DoA), leading to suboptimal results when either of these factors changes.

To address these limitations, the selective fixed-filter ANC (SFANC) method has been proposed to select the most appropriate control filter for various incoming noise types, utilizing either traditional signal processing techniques \cite{shi2020feedforward,shi2019selective} or deep learning-based approaches \cite{luo2022hybrid, wang2025deep,wang2025transferable}. This method provides a more robust solution for noise sources with diverse frequency characteristics, while maintaining low computational cost and fast response time. However, SFANC methods mainly focus on the frequency content of the noise and overlook spatial information, such as DoA, which has also been proven to be critical for the performance of ANC systems \cite{liebich2018direction}.

To date, several researchers have developed methods to incorporate the influence of DoA into ANC systems. Some studies have aimed to enhance noise cancellation from various directions by improving the causality of the ANC system \cite{zhang2014causality}. Others have focused on achieving spatial selectivity, in which unwanted noise is suppressed while the desired sound is preserved. This has been accomplished either by initially canceling both the noise and the desired sound and subsequently reproducing the desired sound \cite{patel2019design,zhang2023directional}, or by selectively controlling the unwanted noise while leaving the desired sound unaffected \cite{xiao2023spatially}. However, these techniques rely on adaptive algorithms for real-time control filter updates. Toyooka et al. proposed a method that selects fixed filters corresponding to different noise source locations \cite{toyooka2025active}. Su et al. proposed an alternative approach that considers both the frequency and directional information of the noise source for control filter selection in a multichannel ANC system \cite{su2024spatial,su2025co}. However, these methods are limited to free-field environments and rely on traditional signal processing techniques for DoA estimation, which are ineffective in handling noise sources under complex reverberant environments. In contrast, deep neural networks (DNNs) are powerful models for accurate source locations without requiring specific modeling assumptions, making them well-suited for such challenging acoustic conditions \cite{li2023doa}.

To address this limitation, this paper proposes a convolutional neural network (CNN)-based directional SFANC method that incorporates DoA information of the noise source in reverberant environments. To achieve this, a CNN trained using multi-task learning is deployed on a co-processor to estimate the elevation and azimuth angles of the noise source based on multiple reference signals, and to select the most appropriate control filter at the frame level. The selected control filter is then applied on a sample-by-sample basis to enable delayless noise control.

The remainder of this paper is organized as follows. Section II introduces the fundamentals of DNN-based DoA estimation and the multi-reference ANC system. Section III describes the overall framework and details of the proposed directional SFANC method. Section IV evaluates the performance of the proposed algorithm through numerical simulations. Finally, Section V concludes the paper with future research directions.

\section{Preliminaries of DNN-based Directional ANC}
\subsection{General Principle of DNN-based DoA Estimation}
DNN-based DoA estimation depends on the multichannel signals recorded with an array of $J$ microphones spatially distributed in the environment, which collectively capture the directional information of the sound source. In reverberant environments, the signal received at the $j$-th microphone can be modeled as
\begin{equation}
{r_j}(n) = {q_j}(n) * x(n), 
\end{equation}
where ${r_j}(n)$ denotes the signal received at the 
$j$-th microphone, $x(n)$ is the source signal, ${q_j}(n)$ represents the room impulse responses (RIRs) between the source and the $j$-th microphone and $*$ denotes the linear convolution operator.

Due to differences in source location, microphone location, and room acoustics, each microphone captures a version of the source signal convolved with a distinct RIR. These variations result in interchannel differences in delay and amplitude, which encode spatial information about the DoA of the sound source relative to the microphone array. DNNs can automatically identify the relationship between the multichannel signal features and the sound source location. This offers a significant advantage over traditional DoA estimation techniques, which often rely on specific modeling assumptions and tend to perform poorly in complex indoor environments. Accordingly, the proposed directional SFANC framework employs a CNN to estimate the DoA, specifically the azimuth angle $\theta$ and elevation angle $\phi$ of the noise source.

\subsection{Multi-Reference Active Noise Control System}
In addition to DoA estimation, the multichannel signals can also serve as reference inputs for the ANC system. Figure 1 illustrates the block diagram of the multi-reference ANC system that consists of $J$ reference microphones, $1$ secondary source and $1$ error microphone. The control signal utilized to drive the secondary source is expressed as
\begin{equation}
y(n) = \sum\limits_{j = 1}^J {{\mathbf{r}}_j^{\rm T}(n)} {{\mathbf{w}}_j}(n),
\end{equation}
where ${{\mathbf{r}}_j}(n) = {[{r_j}(n),{r_j}(n - 1), \cdots ,{r_j}(n - L + 1)]^{\rm{T}}}$ is the $j$-th reference signal vector, ${{\mathbf{w}}_j}(n) = {[{w_{j,1}}(n),{w_{j,2}}(n), \cdots ,{w_{j,L}}(n)]^{\rm{T}}}$ is the $j$-th control filter vector, $L$ is the control filter length and $\rm T$ denotes the matrix transpose operator.

The error signal picked up by the error microphone is stated as 
\begin{equation}
e(n) = d(n) - {{\mathbf{y}}^{\rm T}}(n){\mathbf{s}}(n),
\end{equation}
where $d(n)$ is the disturbance at the error microphone, ${\mathbf{y}}(n) = {[y(n),y(n - 1), \cdots ,y(n - {L_s} + 1)]^{\rm{T}}}$ is the control signal vector, ${\mathbf{s}}(n) = {[{s_1}(n),{s_2}(n), \cdots ,{s_{{L_s}}}(n)]^{\rm{T}}}$ is the secondary path impulse response and $L_s$ is the secondary path length. 

According to the FxLMS algorithm, the $j$-th control filter coefficient is updated by
\begin{equation}
{{\mathbf{w}}_j}(n + 1) = {{\mathbf{w}}_j}(n) + \mu {{{\mathbf{r}}}_j'}(n)e(n),
\end{equation}
where $\mu$ is the stepsize, ${{{\mathbf{r}}}_j'}(n)$ is the filtered reference signal generated by passing the reference signal through the estimated secondary path as 
\begin{equation}
{{{\mathbf{r}}}_j'}(n) = \hat s(n) * {{\mathbf{r}}_j}(n)
\end{equation}

However, the adaptive FxLMS algorithm requires time to converge and an inappropriate step size may even lead to noise amplification. To improve response time and enhance system stability, the proposed directional SFANC method selects a pre-trained fixed-filter based on the DoA of the noise source.

\begin{figure}[t]
\begin{center}
\includegraphics[width=90mm]{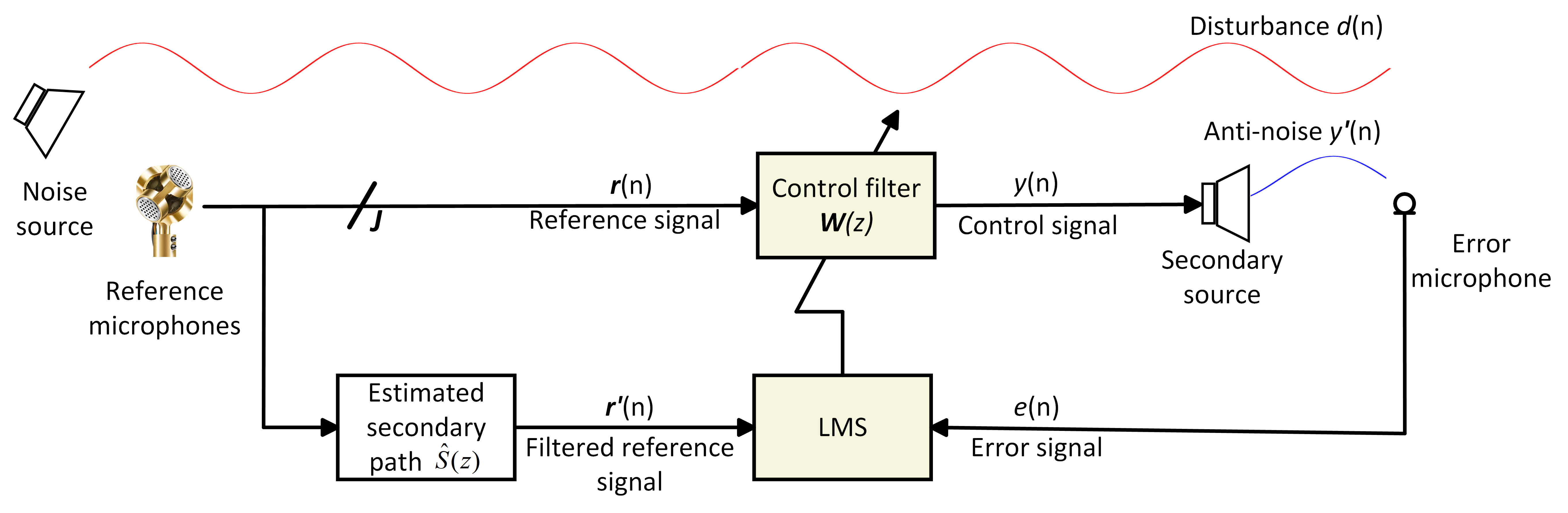}
\end{center}
\caption{Block diagram of the multi-reference active noise control system.}
\vspace*{-3pt}
\end{figure}

\section{Proposed Directional SFANC Method}
The block diagram of the proposed directional SFANC technique is shown in Fig. 2. The system consists of two main components: a real-time controller for noise cancellation and a co-processor for control filter selection. As illustrated in Fig. 2, the controller operates at the sampling rate to generate the control signal. Meanwhile, reference signals are collected at each frame and transmitted to the co-processor. The objective is to estimate the azimuth and elevation angles of the noise source relative to the reference microphone array, using information embedded in the reference signals. Based on these inputs, the CNN outputs the predicted probabilities for each azimuth and elevation class as
\begin{equation}
({{{\mathbf{\hat p}}}}_{\rm{azim}},{{{\mathbf{\hat p}}}}_{\rm{elev}}) = CNN({{\mathbf{R}}};{\Theta ^*})
\end{equation}
where ${{\mathbf{R}}}$ is the short-term fourier transform (STFT) spectrograms of $J$-channel reference signals at one frame, ${\Theta ^*}$ is the trained CNN parameters, ${{{\mathbf{\hat p}}}}_{\rm{azim}} = [{{\hat p}}_{\rm{azim},1},{{\hat p}}_{\rm{azim},2},...,{{\hat p}}_{\rm{azim},A}]$ is the predicted probability distribution over the $A$ azimuth angle classes, ${{{\mathbf{\hat p}}}}_{\rm{elev}} = [{{\hat p}}_{\rm{elev},1},{{\hat p}}_{\rm{elev},2},...,{{\hat p}}_{\rm{elev},B}]$ is the predicted probability distribution over the $B$ elevation angle classes.

Then, the estimated azimuth index $\hat a$ can be obtained as
\begin{equation}
\hat a = \mathop {\arg \max }\limits_{i \in \{ 1,2,...,A\} } {{\hat p}_{\rm{azim},i}}.
\end{equation}
where ${{\hat p}_{\rm{azim},i}}$ denotes the predicted probability of the $i$-th azimuth class.

And the estimated elevation index $\hat b$ can be obtained as
\begin{equation}
\hat b = \mathop {\arg \max }\limits_{k \in \{ 1,2,...,B\} } {{\hat p}_{\rm{elev},k}}.
\end{equation}
where ${{\hat p}_{\rm{elev},k}}$ denotes the predicted probability of the $k$-th elevation class.

Finally, based on the predicted azimuth index and elevation index, the coefficients of the selected control filter are updated at the frame rate. By facilitating cooperation between the real-time noise controller and the co-processor, the proposed directional SFANC method enables delayless noise control.

\begin{figure}[t]
\begin{center}
\includegraphics[width=90mm]{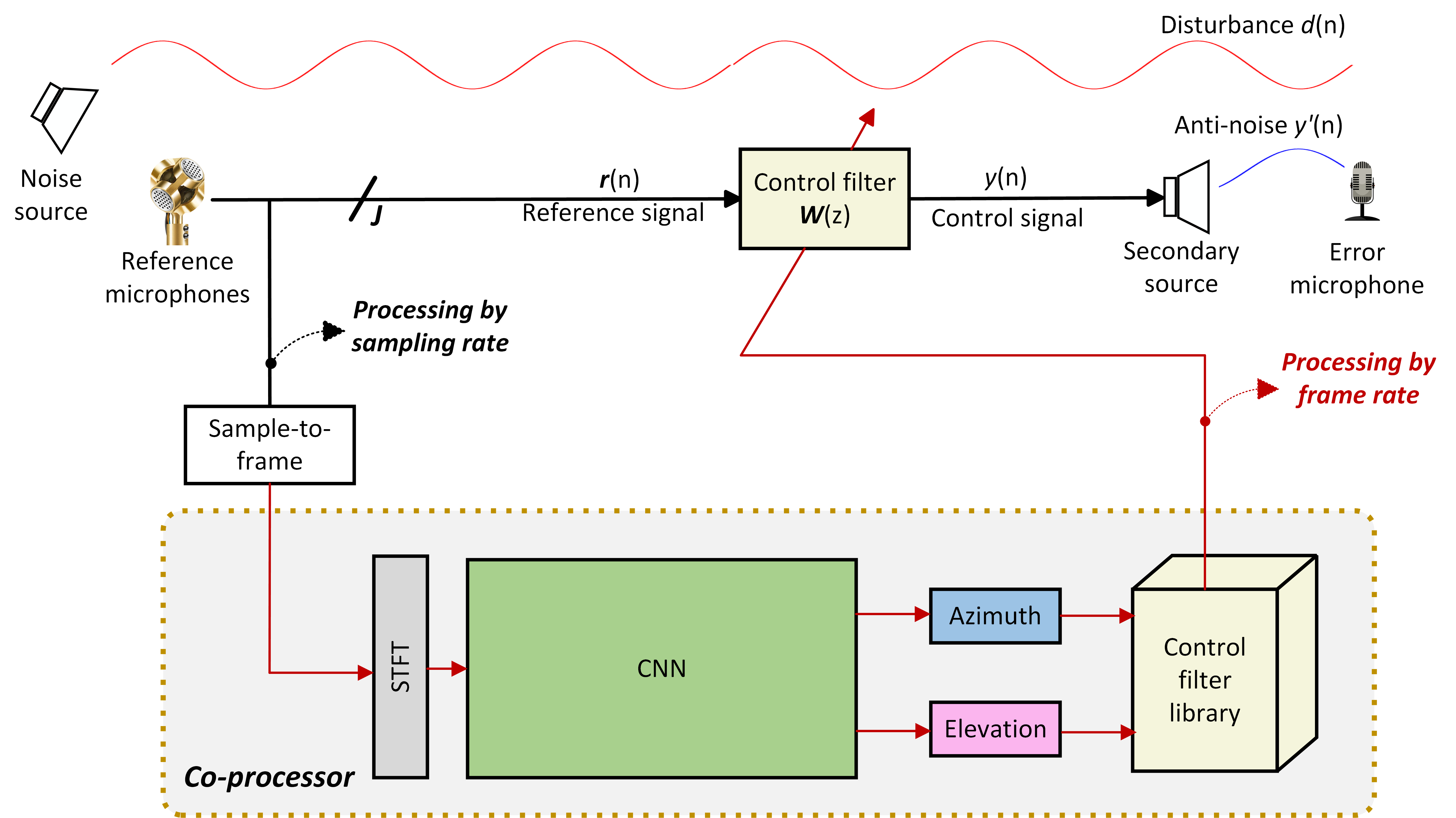}
\end{center}
\caption{Block diagram of the proposed directional SFANC method.}
\vspace*{-3pt}
\end{figure}

\subsection{Pre-trained Control Filter Library}
Prior to the online execution of the directional SFANC method, a control filter library must be pre-trained to account for various combinations of azimuth and elevation angles of the noise source. Specifically, the horizontal plane of the reference microphone array is divided into $A$ azimuth classes, and the vertical plane is divided into $B$ elevation classes. At each discrete direction defined by this grid, the FxLMS algorithm is used to pre-train a control filter for broadband noise. These control filters are then stored in a library for deployment.

\subsection{CNN Trained Using a Multi-Task Learning Strategy}
In this work, a CNN is employed for DoA estimation, which has been found to be effective for this task \cite{grumiaux2022survey}. The architecture of the proposed CNN is illustrated in Fig. 3. The input consists of a one-frame, $J$-channel reference signal, which is transformed into $J$ magnitude spectrograms and $J$ phase spectrograms using the STFT for feature extraction. The pre-processed data is passed through three convolutional modules, each comprising a convolutional layer followed by group normalization, ReLU activation, and max-pooling. Adaptive average pooling is then applied to reduce the feature maps by averaging over both frequency and time. These pooled features are subsequently fed into two fully connected (FC) layers to estimate the class probabilities for $A$ azimuth and $B$ elevation angles. Final predictions are obtained through softmax layers.

\begin{figure}[t]
\begin{center}
\includegraphics[width=30mm]{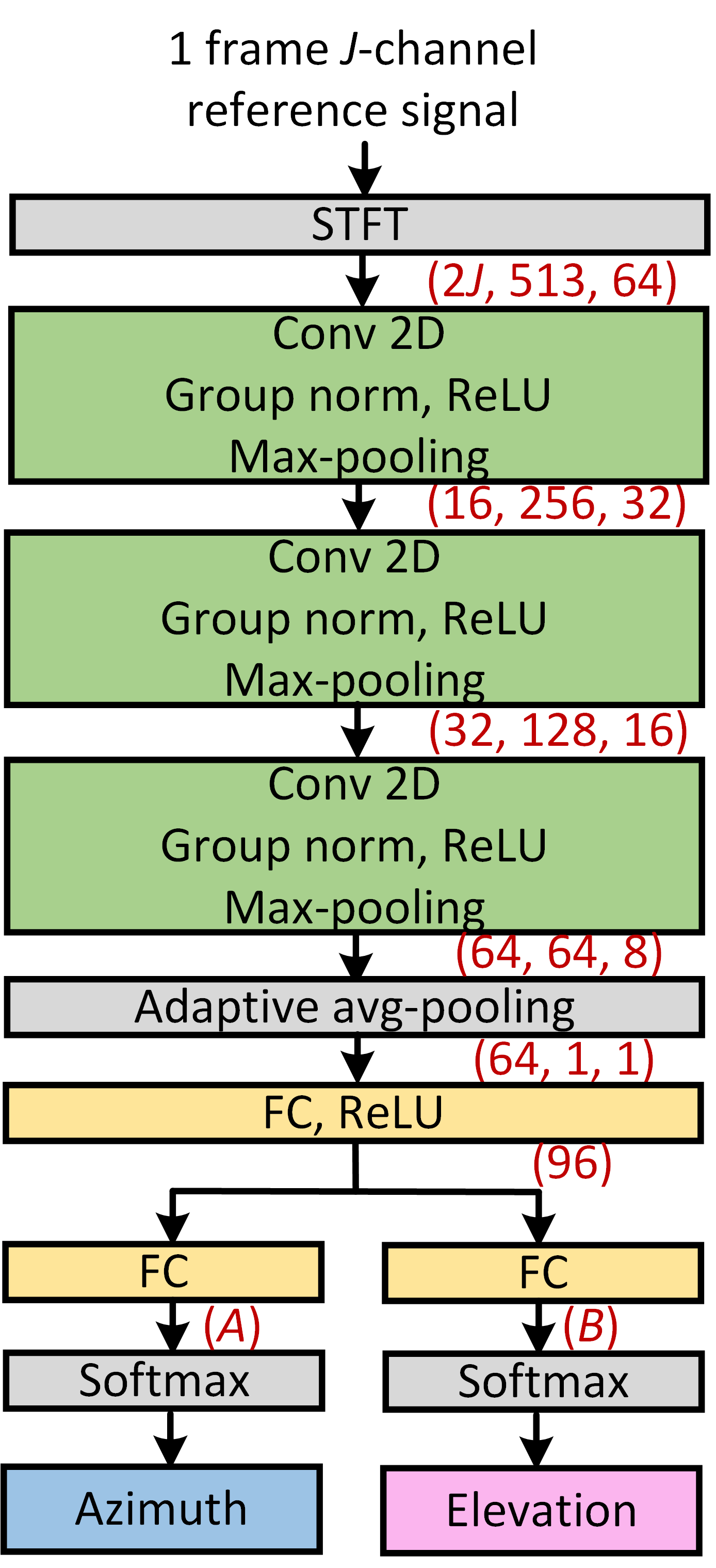}
\end{center}
\caption{Architecture of the proposed CNN.}
\vspace*{-3pt}
\end{figure}

As previously discussed, the selected control filter is determined by the azimuth and elevation angle indices predicted by the CNN. To enable this, the CNN is trained using a multi-task learning strategy that simultaneously performs azimuth and elevation classification. The loss functions for both tasks are cross-entropy loss functions, denoted as $Loss{_{\rm{azim}}}$ and $Loss{_{\rm{elev}}}$ respectively. The joint loss function used to train the CNN is formulated as 
\begin{equation}
Loss = Loss{_{\rm{azim}}} + Loss{_{\rm{elev}}}
\end{equation}

This joint loss allows the network to learn shared representations while balancing the learning of both tasks, offering a more efficient alternative to training separate models \cite{zhang2021survey}.

\section{Numerical Simulations}
The effectiveness of the proposed directional SFANC method is evaluated in a 4×1×1 multi-reference ANC system in reverberant environments, where a four-channel tetrahedral microphone array with a diameter of 2.5 cm is employed as the reference microphone array to effectively capture the spatial characteristics of the noise source \cite{kushwaha2023sound}. For the construction of the pre-trained control filter library, as illustrated in Fig. 4, the horizontal plane of the microphone array is divided into six azimuth classes: 0°, 60°, 120°, 180°, 240°, and 300°, while the vertical plane is divided into three elevation classes: 90°, 30°, and -30°. The distance between the noise source and the microphone array is fixed at 0.2 m. At each discrete direction defined by this spatial grid, a control filter is pre-trained using the FxLMS algorithm with broadband noise in the 20–2020 Hz range, targeting the low-frequency band typically addressed by ANC systems. A total of 13 control filters are trained and stored in the library. The STFT is computed using a Hann window of 1024 samples, a hop size of 64 samples, and the system operates at a sampling frequency of 16 kHz.

\begin{figure}[t]
\begin{center}
\includegraphics[width=85mm]{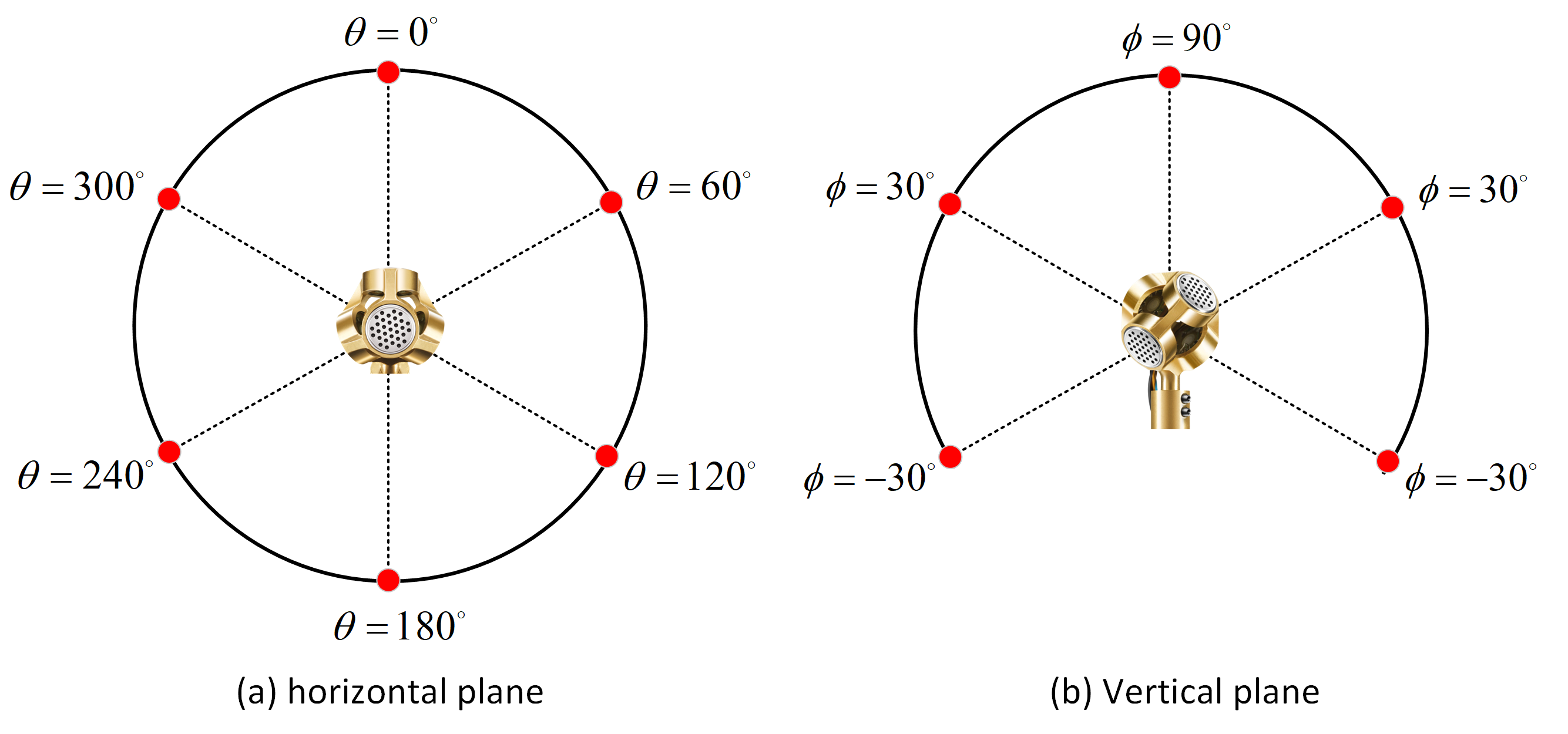}
\end{center}
\caption{Illustration of the pre-trained control filter library: (a) azimuth angle classes defined in the horizontal plane, and (b) elevation angle classes defined in the vertical plane.}
\vspace*{-3pt}
\end{figure}

\begin{table}[h!]
    \centering
    \caption{Configurations for training, validation and testing datasets.}
    \begin{tabular}{|c|c|}
        \hline
        \multicolumn{2}{|c|}{\textbf{Training and Validation Datasets}} \\
        \hline
        Noise signal & Synthesized noises $\&$ real noises\\
        \hline
        Room size (m) & R1: (6x4x3); R2: (12x8x3.5); R3: (16x14x4)\\
        \hline
        Array positions & 8 arbitrary positions in each room \\
        \hline
        RT$_{60}$ (s) & R1: 0.1, 0.2, 0.3; R2: 0.4, 0.5, 0.6; R3: 0.7, 0.8, 0.9 \\
        \hline
        SNR (dB) & Uniformly sampled from 30 to 50 \\
        \hline
        \multicolumn{2}{|c|}{\textbf{Testing Dataset}} \\
        \hline
        Noise signal & Synthesized noises $\&$ real noises\\
        \hline
        Room size (m) & (11x9x3.2)\\
        \hline
        Array positions & 4 arbitrary positions in each room \\
        \hline
        RT$_{60}$ (s) & 0.48 \\
        \hline
        SNR (dB) & 30, 40, 50 \\
        \hline
    \end{tabular}
\end{table}

\begin{table}[h!]
    \centering
    \caption{The classification accuracy of the CNN with different SNRs.}
    \begin{tabular}{|c|c|c|c|}
        \hline
        \textbf{Metrics} 
        & \textbf{SNR = 30 dB} 
        & \textbf{SNR = 40 dB} 
        & \textbf{SNR = 50 dB} \\
        \hline
        Azimuth angle Acc.
        & 96.4\% & 96.4\% & 96.4\% \\
        \hline
        Elevation angle Acc.
        & 90.7\% & 90.8\% & 91.0\% \\
        \hline
    \end{tabular}
\end{table}

\subsection{DoA Estimation with CNN}
\subsubsection{Dataset generation}
A noise dataset consisting of both synthetic and real noise signals is first constructed. The synthetic noises are generated as bandlimited white noise with varying bandwidths, while the real noises are sourced from the UrbanSound8K dataset \cite{salamon2014dataset}. The generated noise signals are then convolved with a variety of RIRs to produce reference signals captured by the tetrahedral microphone array, with additive noise included. The azimuth angle of the noise source relative to the microphone array is randomly selected from the range [0°, 360°], the elevation angle from [-60°, 90°], and the source-to-array distance from [0.1 m, 0.6 m], with the array-to-surface distance maintained at over 1 m. The nearest azimuth and elevation angle classes, as defined in Fig. 4, are assigned as the corresponding labels. To enhance the system's robustness under adverse acoustic conditions, the training and validation datasets incorporate variations of the RIRs made with different room sizes, array positions, reverberation time (RT$_{60}$), and signal-to-noise ratio (SNR) levels. For the testing dataset, both the noise types and acoustic environments are distinct from those used during training, ensuring a fair evaluation of generalization performance. A summary of the dataset configuration is provided in Table I. The RIRs are generated using the \textit{gpuRIR} library \cite{diaz2021gpurir} based on the image method \cite{allen1979image}. 

In total, the dataset includes 46080 training samples (38400 synthetic, 7680 real), 5760 validation samples (4800 synthetic, 960 real), and 4800 test samples (4000 synthetic, 800 real). Each sample is a four-channel, 0.5-second frame.

\subsubsection{Classification accuracy under unseen acoustic environments and noise types}
As shown in Table II, the proposed CNN achieves a classification accuracy of approximately 96\% for azimuth angle and 91\% for elevation angle under varying SNR levels. Notably, these results are obtained using unseen noise types and acoustic environments during testing, highlighting the model's strong generalization capability and robustness to real-world variability. In addition, the CNN is computationally efficient, with only 0.03 million parameters, a CPU runtime of 7.83 ms, and 119.86 million multiply-accumulate operations (MACs), making it suitable for deployment on embedded devices. These findings demonstrate the effectiveness of the proposed method in selecting appropriate control filters for noise sources with different DoAs, thereby enabling efficient noise control in reverberant environments.

\subsection{Noise Cancellation based on DoA Estimation}
Noise cancellation simulations are conducted in the testing environments summarized in Table 1 with the arrangement of the 4x1x1 multi-reference ANC system shown in Fig. 5. The proposed directional SFANC method is compared with several baseline methods, including the conventional FxLMS algorithm, standard SFANC \cite{luo2024real}, and generative fixed-filter ANC (GFANC) \cite{luo2025mssp}. For both the standard SFANC and GFANC methods, the pre-trained control filters are trained using a noise source located at $(\theta = 0^\circ, \phi = 30^\circ)$, corresponding to the center of the reference microphone array.

\begin{figure}[t]
\begin{center}
\includegraphics[width=85mm]{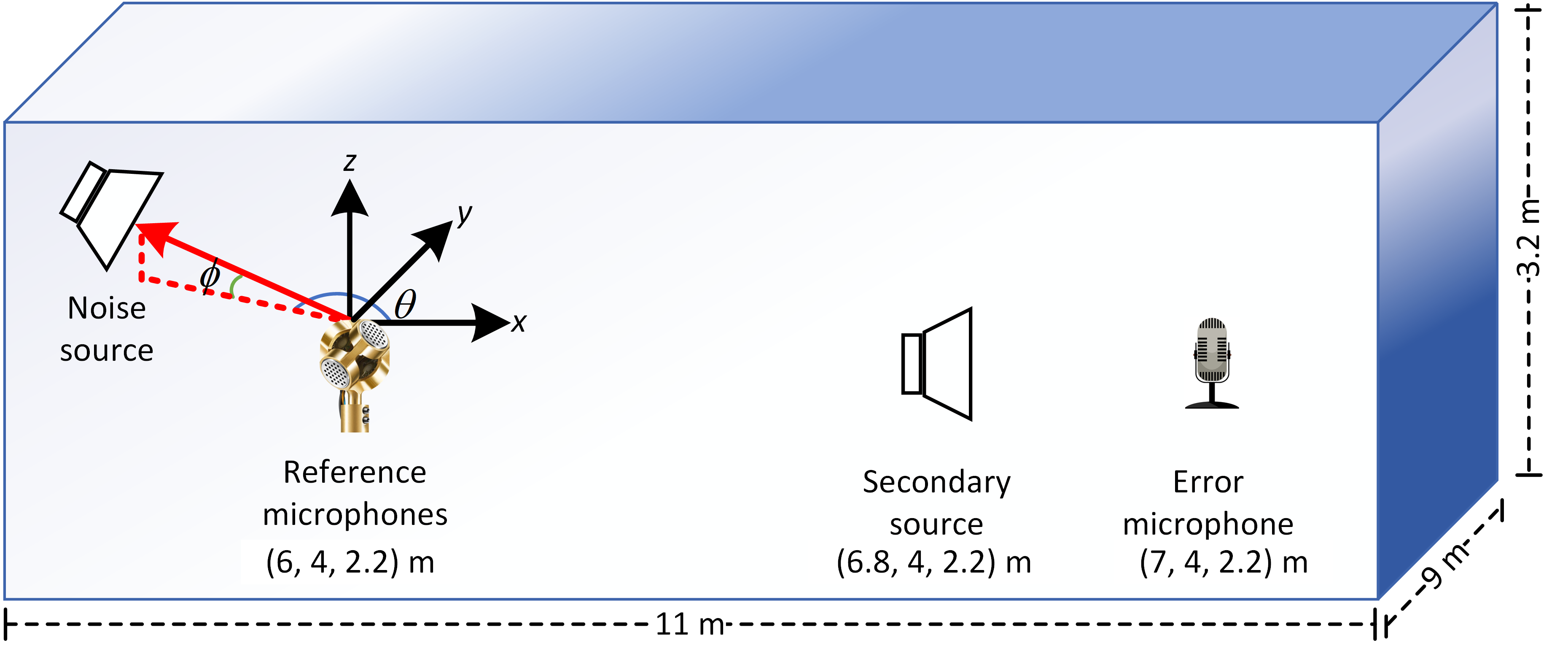}
\end{center}
\caption{Simulation arrangement of the 4x1x1 multi-reference ANC system.}
\vspace*{-3pt}
\end{figure}

\subsubsection{Broadband Noise Cancellation}
To evaluate the noise reduction performance of the proposed directional SFANC method for broadband noise, a primary noise signal in the 100–700 Hz range is first positioned at $(\theta = 120^\circ, \phi = 30^\circ)$ corresponding to the reference microphone array. The power spectral density (PSD) of the residual noise after applying the FxLMS, SFANC, GFANC, and directional SFANC methods is shown in Fig. 6(a). The PSD is computed by averaging across the entire duration of the signal and the step size of the FxLMS algorithm is set to $1 \times {10^{ - 4}}$. Additionally, the averaged noise reduction levels per 0.5 second achieved by the four methods are presented in Fig. 6(b). A similar evaluation is conducted for a noise source located at $(\theta = 0^\circ, \phi = -30^\circ)$ corresponding to the reference microphone array, with the results shown in Fig. 7. As shown in Fig. 6 and Fig. 7, the proposed directional SFANC method outperforms the conventional FxLMS algorithm in terms of both response time and noise reduction performance. Furthermore, when the azimuth or elevation angle of the noise source deviates from the pre-trained filter location used in the SFANC and GFANC methods, these approaches exhibit limited noise reduction capability or may even amplify the noise. This degradation is attributed to their inability to track changes in the DoA of the noise source.

\begin{figure}[t]
\begin{center}
\includegraphics[width=90mm]{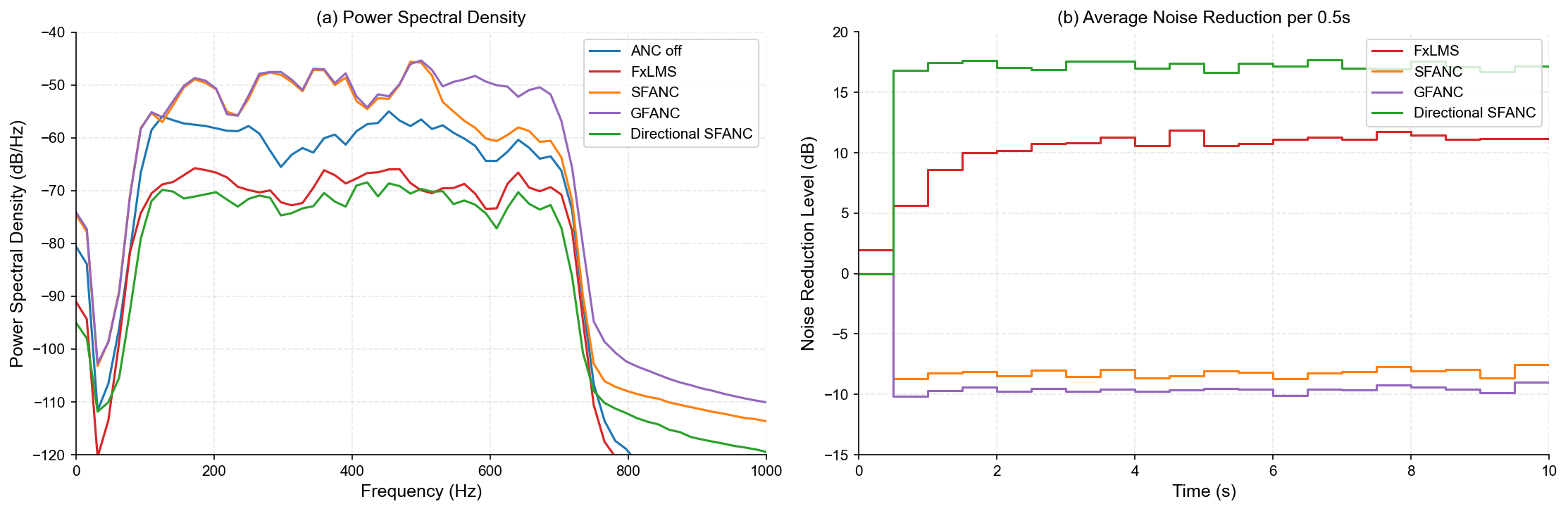}
\end{center}
\caption{(a) PSD and (b) averaged noise reduction level per 0.5 second attenuated by different ANC algorithms for 100–700 Hz broadband noise located at $(\theta = 120^\circ, \phi = 30^\circ)$ relative to the reference microphone array.}
\vspace*{-3pt}
\end{figure}

\begin{figure}[t]
\begin{center}
\includegraphics[width=90mm]{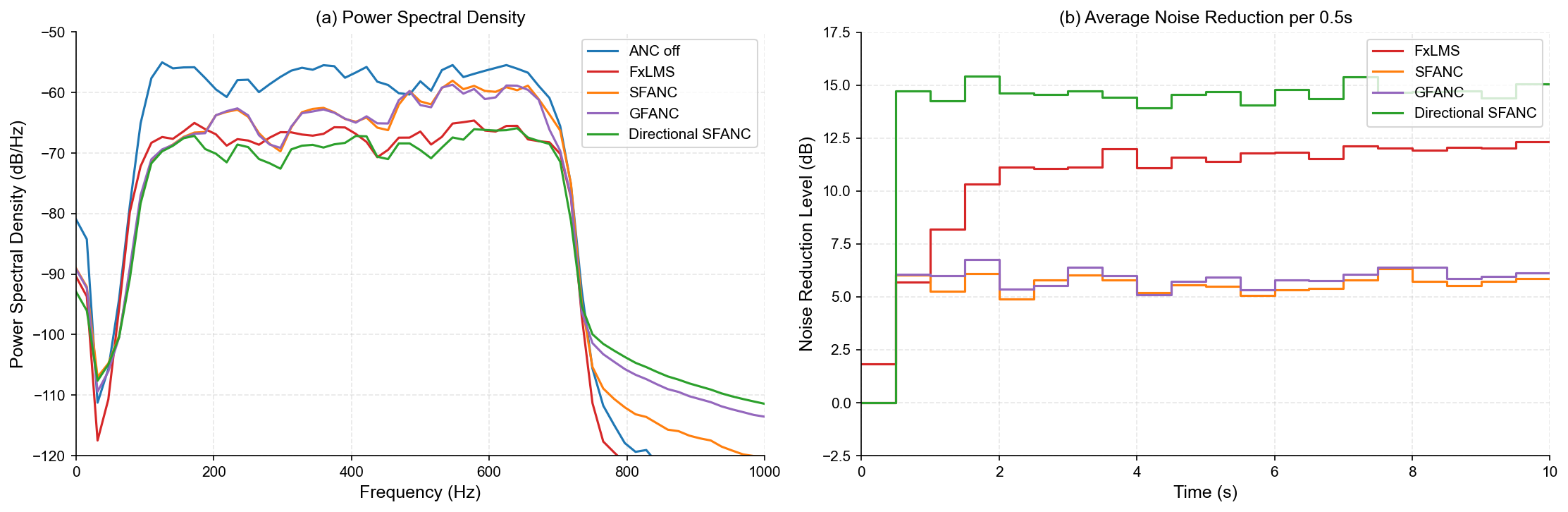}
\end{center}
\caption{(a) PSD and (b) averaged noise reduction level per 0.5 second attenuated by different ANC algorithms for 100–700 Hz broadband noise located at $(\theta = 0^\circ, \phi = -30^\circ)$ relative to the reference microphone array.}
\vspace*{-3pt}
\end{figure}

\subsubsection{Real-world Noise Cancellation}
To evaluate the noise reduction performance of the proposed directional SFANC method on real-world noise, a washing machine noise source is positioned at $(\theta = 110^\circ, \phi = -15^\circ)$ corresponding to the reference microphone array. The PSD of the residual noise after applying the FxLMS, SFANC, GFANC, and directional SFANC methods is shown in Fig. 8(a). In addition, the average noise reduction levels per 0.5-second interval achieved by the four methods are presented in Fig. 8(b). It can be observed that the proposed directional SFANC method continues to achieve superior noise reduction performance and faster response time compared to the FxLMS algorithm, even when the noise source is located at a position not included in the pre-trained control filter library. In contrast, the SFANC and GFANC methods tend to amplify the noise due to their inability to adapt to variations in the DoA of the noise source.

\begin{figure}[t]
\begin{center}
\includegraphics[width=90mm]{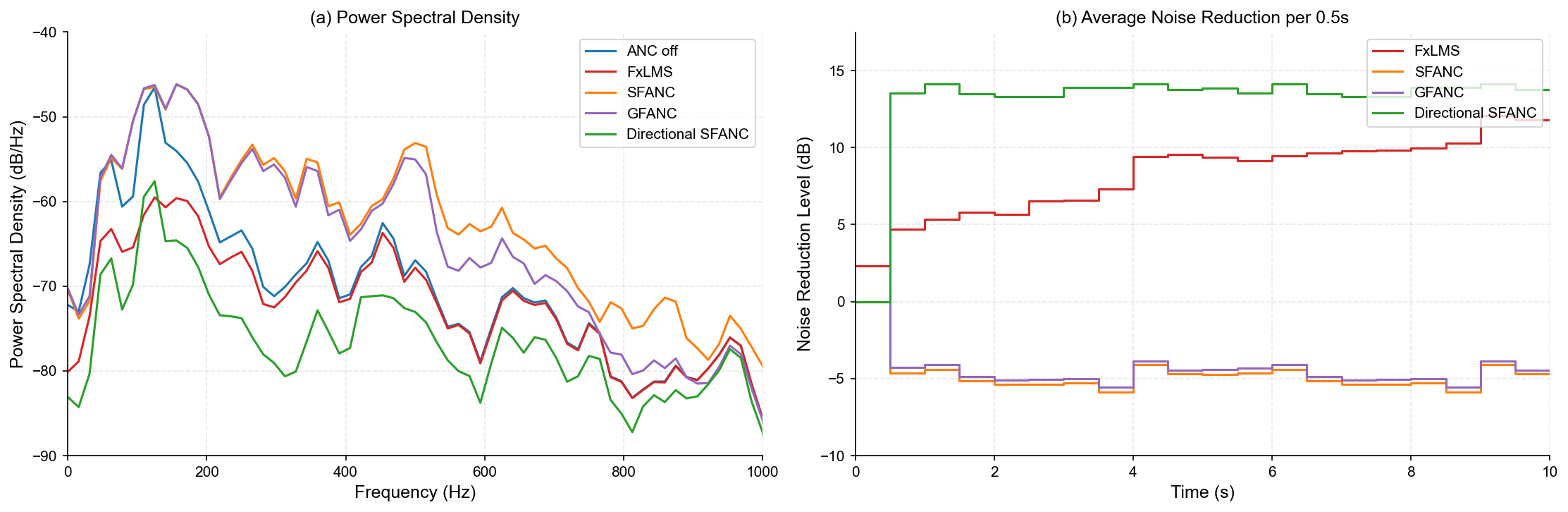}
\end{center}
\caption{(a) PSD and (b) averaged noise reduction level per 0.5 second attenuated by different ANC algorithms for washing machine noise located at $(\theta = 110^\circ, \phi = -15^\circ)$ relative to the reference microphone array.}
\vspace*{-3pt}
\end{figure}

\section{Conclusions}
This paper proposes a novel directional SFANC method to tackle noise sources with varying DoA in complex reverberant environments. A lightweight CNN, trained via multi-task learning, is employed to dynamically select the optimal control filter based on the reference signals. Simulation results show that the proposed method significantly outperforms the conventional FxLMS algorithm for both broadband and real-world noises across various DoAs, offering faster response and superior noise reduction under reverberant conditions. In contrast, many existing learning-based ANC approaches struggle to adapt to directional variations in the noise source.

Despite its advantages, the current method does not consider the source-to-array distance, which may also affect ANC performance. Future work will address this by integrating distance estimation into the control framework.


\printbibliography

@article{xiao2023spatially,
  title={Spatially selective active noise control systems},
  author={Xiao, Tong and Xu, Buye and Zhao, Chuming},
  journal={The Journal of the Acoustical Society of America},
  volume={153},
  number={5},
  pages={2733--2733},
  year={2023},
  publisher={AIP Publishing}
}

@article{elliott1993active,
  title={Active noise control},
  author={Elliott, Stephen J and Nelson, Philip Arthur},
  journal={IEEE signal processing magazine},
  volume={10},
  number={4},
  pages={12--35},
  year={1993},
  publisher={IEEE}
}

@article{kuo1999active,
  title={Active noise control: a tutorial review},
  author={Kuo, Sen M and Morgan, Dennis R},
  journal={Proceedings of the IEEE},
  volume={87},
  number={6},
  pages={943--973},
  year={1999},
  publisher={IEEE}
}

@article{morgan1980analysis,
  title={An analysis of multiple correlation cancellation loops with a filter in the auxiliary path},
  author={Morgan, D},
  journal={IEEE Transactions on Acoustics, Speech, and Signal Processing},
  volume={28},
  number={4},
  pages={454--467},
  year={1980},
  publisher={IEEE}
}

@article{shen2022adaptive,
  title={Adaptive-gain algorithm on the fixed filters applied for active noise control headphone},
  author={Shen, Xiaoyi and Shi, Dongyuan and Gan, Woon-Seng and Peksi, Santi},
  journal={Mechanical Systems and Signal Processing},
  volume={169},
  pages={108641},
  year={2022},
  publisher={Elsevier}
}

@article{jung2019local,
  title={Local active control of road noise inside a vehicle},
  author={Jung, Woomin and Elliott, Stephen J and Cheer, Jordan},
  journal={Mechanical Systems and Signal Processing},
  volume={121},
  pages={144--157},
  year={2019},
  publisher={Elsevier}
}

@article{shi2020feedforward,
  title={Feedforward selective fixed-filter active noise control: Algorithm and implementation},
  author={Shi, Dongyuan and Gan, Woon-Seng and Lam, Bhan and Wen, Shulin},
  journal={IEEE/ACM Transactions on Audio, Speech, and Language Processing},
  volume={28},
  pages={1479--1492},
  year={2020},
  publisher={IEEE}
}

@article{luo2024real,
  title={Real-time implementation and explainable AI analysis of delayless CNN-based selective fixed-filter active noise control},
  author={Luo, Zhengding and Shi, Dongyuan and Ji, Junwei and Shen, Xiaoyi and Gan, Woon-Seng},
  journal={Mechanical Systems and Signal Processing},
  volume={214},
  pages={111364},
  year={2024},
  publisher={Elsevier}
}

@inproceedings{wang2025transferable,
  title={Transferable Selective Virtual Sensing Active Noise Control Technique Based on Metric Learning},
  author={Wang, Boxiang and Shi, Dongyuan and Luo, Zhengding and Shen, Xiaoyi and Ji, Junwei and Gan, Woon-Seng},
  booktitle={ICASSP 2025-2025 IEEE International Conference on Acoustics, Speech and Signal Processing (ICASSP)},
  pages={1--5},
  year={2025},
  organization={IEEE}
}

@inproceedings{ji2025preventing,
  title={Preventing output saturation in active noise control: An output-constrained Kalman filter approach},
  author={Ji, Junwei and Shi, Dongyuan and Wang, Boxiang and Shen, Xiaoyi and Luo, Zhengding and Gan, Woon-Seng},
  booktitle={ICASSP 2025-2025 IEEE International Conference on Acoustics, Speech and Signal Processing (ICASSP)},
  pages={1--5},
  year={2025},
  organization={IEEE}
}

@inproceedings{shi2019selective,
  title={Selective virtual sensing technique for multi-channel feedforward active noise control systems},
  author={Shi, Chuang and Xie, Rong and Jiang, Nan and Li, Huiyong and Kajikawa, Yoshinobu},
  booktitle={ICASSP 2019-2019 IEEE International Conference on Acoustics, Speech and Signal Processing (ICASSP)},
  pages={8489--8493},
  year={2019},
  organization={IEEE}
}

@article{zhang2023directional,
  title={A directional spatial active noise control system with a sound field separation algorithm},
  author={Zhang, Huawei and Zhang, Jihui and Ma, Fei and Sun, Huiyuan and Samarasinghe, Prasanga N},
  journal={The Journal of the Acoustical Society of America},
  volume={154},
  number={4\_supplement},
  pages={A162--A162},
  year={2023},
  publisher={Acoustical Society of America}
}

@article{su2025co,
  title={Co-forecasting of Time-varying Spatial-frequency Map for Selective Fixed-Filter Multichannel ANC based on Dynamic Factor Graph},
  author={Su, Xiruo and Shi, Dongyuan and Wu, Bin and Ye, Lingyun and Gan, Woon-Seng},
  journal={IEEE Transactions on Audio, Speech and Language Processing},
  year={2025},
  publisher={IEEE}
}

@article{su2024spatial,
  title={Spatial-Frequency-Based Selective Fixed-Filter Algorithm for Multichannel Active Noise Control},
  author={Su, Xiruo and Shi, Dongyuan and Zhu, Zhijuan and Gan, Woon-Seng and Ye, Lingyun},
  journal={IEEE Signal Processing Letters},
  year={2024},
  publisher={IEEE}
}

@inproceedings{liebich2018direction,
  title={Direction-of-arrival dependency of active noise cancellation headphones},
  author={Liebich, Stefan and Richter, Jan-Gerrit and Fabry, Johannes and Durand, Christopher and Fels, Janina and Jax, Peter},
  booktitle={Noise Control and Acoustics Division Conference},
  volume={51425},
  pages={V001T08A003},
  year={2018},
  organization={American Society of Mechanical Engineers}
}

@article{zhang2014causality,
  title={Causality study on a feedforward active noise control headset with different noise coming directions in free field},
  author={Zhang, Limin and Qiu, Xiaojun},
  journal={Applied Acoustics},
  volume={80},
  pages={36--44},
  year={2014},
  publisher={Elsevier}
}

@article{patel2019design,
  title={Design and implementation of an active noise control headphone with directional hear-through capability},
  author={Patel, Vinal and Cheer, Jordan and Fontana, Simone},
  journal={IEEE Transactions on Consumer Electronics},
  volume={66},
  number={1},
  pages={32--40},
  year={2019},
  publisher={IEEE}
}

@article{toyooka2025active,
  title={Active Noise Control Systems with Sound Source Localization Robust to Noise Source Movement},
  author={Toyooka, Shota and Kajikawa, Yoshinobu},
  journal={IEICE Transactions on Fundamentals of Electronics, Communications and Computer Sciences},
  volume={108},
  number={2},
  pages={160--164},
  year={2025},
  publisher={The Institute of Electronics, Information and Communication Engineers}
}

@article{grumiaux2022survey,
  title={A survey of sound source localization with deep learning methods},
  author={Grumiaux, Pierre-Amaury and Kiti{\'c}, Sr{\dj}an and Girin, Laurent and Gu{\'e}rin, Alexandre},
  journal={The Journal of the Acoustical Society of America},
  volume={152},
  number={1},
  pages={107--151},
  year={2022},
  publisher={AIP Publishing}
}

@inproceedings{li2023doa,
  title={DoA Estimation of Room Reflections Using NN-Based MUSIC Algorithm},
  author={Li, Haowen and Zhang, Wen and Zhang, Lijun},
  booktitle={2023 Asia Pacific Signal and Information Processing Association Annual Summit and Conference (APSIPA ASC)},
  pages={1960--1965},
  year={2023},
  organization={IEEE}
}

@article{luo2022hybrid,
  title={A hybrid sfanc-fxnlms algorithm for active noise control based on deep learning},
  author={Luo, Zhengding and Shi, Dongyuan and Gan, Woon-Seng},
  journal={IEEE Signal Processing Letters},
  volume={29},
  pages={1102--1106},
  year={2022},
  publisher={IEEE}
}

@article{diaz2021gpurir,
  title={gpuRIR: A python library for room impulse response simulation with GPU acceleration},
  author={Diaz-Guerra, David and Miguel, Antonio and Beltran, Jose R},
  journal={Multimedia Tools and Applications},
  volume={80},
  number={4},
  pages={5653--5671},
  year={2021},
  publisher={Springer}
}

@article{allen1979image,
  title={Image method for efficiently simulating small-room acoustics},
  author={Allen, Jont B and Berkley, David A},
  journal={The Journal of the Acoustical Society of America},
  volume={65},
  number={4},
  pages={943--950},
  year={1979},
  publisher={Acoustical Society of America}
}

@article{zhang2021survey,
  title={A survey on multi-task learning},
  author={Zhang, Yu and Yang, Qiang},
  journal={IEEE transactions on knowledge and data engineering},
  volume={34},
  number={12},
  pages={5586--5609},
  year={2021},
  publisher={IEEE}
}

@inproceedings{kushwaha2023sound,
  title={Sound source distance estimation in diverse and dynamic acoustic conditions},
  author={Kushwaha, Saksham Singh and Roman, Iran R and Fuentes, Magdalena and Bello, Juan Pablo},
  booktitle={2023 IEEE Workshop on Applications of Signal Processing to Audio and Acoustics (WASPAA)},
  pages={1--5},
  year={2023},
  organization={IEEE}
}

@inproceedings{salamon2014dataset,
  title={A dataset and taxonomy for urban sound research},
  author={Salamon, Justin and Jacoby, Christopher and Bello, Juan Pablo},
  booktitle={Proceedings of the 22nd ACM international conference on Multimedia},
  pages={1041--1044},
  year={2014}
}

@inproceedings{wang2024computation,
  title={Computation-efficient virtual sensing approach with multichannel adjoint least mean square algorithm},
  author={Wang, Boxiang and Ji, Junwei and Shen, Xiaoyi and Shi, Dongyuan and Gan, Woon-Seng},
  booktitle={INTER-NOISE and NOISE-CON Congress and Conference Proceedings},
  volume={270},
  number={10},
  pages={1638--1650},
  year={2024},
  organization={Institute of Noise Control Engineering}
}

@inproceedings{wang2025deep,
  title={DEEP LEARNING-BASED ACTIVE TRIM PANELS FOR ENHANCED AIRCRAFT INTERIOR NOISE CONTROL},
  author={Wang, Boxiang and Misol, Malte and Luo, Zhengding and Ji, Junwei and Shen, Xiaoyi and Shi, Dongyuan and Gan, Woon-Seng},
  booktitle={Proceedings of the 31st International Congress on Sound and Vibration},
  year={2025},
  organization={The Korean Society for Noise and Vibration Engineering}
}

@article{luo2025mssp,
  title={Deep learning-based Generative Fixed-Filter Active Noise Control: Transferability and implementation},
  author={Luo, Zhengding and Ji, Junwei and Wang, Boxiang and Shi, Dongyuan and Ma, Haozhe and Gan, Woon-Seng},
  journal={Mechanical Systems and Signal Processing},
  volume={238},
  pages={113207},
  year={2025},
  publisher={Elsevier}
}

@article{ji2025self,
  title={Self-Boosted Weight-Constrained FxLMS: A Robustness Distributed Active Noise Control Algorithm Without Internode Communication},
  author={Ji, Junwei and Shi, Dongyuan and Luo, Zhengding and Wang, Boxiang and Gan, Woon-Seng},
  journal={IEEE Signal Processing Letters},
  year={2025},
  publisher={IEEE}
}

\end{document}